# Do you know "saudade"? The importance of a cultural and language-based emotion approach for HCI

Diogo Cortiz
Brazilian Network Information Center (NIC.br)
Pontifical Catholic University of São Paulo (PUC-SP)
*São Paulo, Brazil*
diogocortiz@gmail.com

Paulo Boggio
Mackenzie University
São Paulo, Brazil
paulo.boggio@mackenzie.br

*Abstract*— Today most technologies and interfaces are designed to be global. We argue that if emotional aspects are incorporated during the design phase of technologies and interfaces - or in technologies for recognizing users' emotions such as affective computing - culture and language should be taken as relevant criteria.

## I. Introduction

One question that remains open is whether there are basic and universal emotions. In this paper, we review the main theories of emotions, such as Basic Emotion Theory (BET), appraisal theory and constructivist theory. We argue that while there are universal emotional responses, there is room for variation in the interpretation and experience of emotions, often influenced by culture and language. In this sense, we assume the need to think of an emotional approach based on culture and language for HCI.

Today most technologies and interfaces are designed to be global. Big tech companies usually design technologies and interface to be deliverable worldwide. They develop their products in one specific region, usually in the global North, and make them available in markets on different continents, countries, and cultures. Even local startups plan to have a broader market reach if their local technologies. It is a one size fits all approach.

Sometimes those customize their technologies to attend to the specific local demands, most of the time related to policies or local legislation. We argue that if emotional aspects are incorporated during the design phase of technologies and interfaces - or in technologies for recognizing users' emotions such as affective computing - culture and language should be taken as relevant criteria.

## II. Cultural and language-based emotion approach for HCI

### A. Emotion Theories

Today we can divide the studies of emotions into a few approaches. The Basic Emotion Theory (BET) and the constructivist theory are the two main approaches, while Appraisal Theories often are placed between them.
The Basic Emotion Theory (BET) argues that emotions are discrete with clear and well-defined boundaries, biologically fixed, and they are universal for all human beings [7] [8] [11]. This theory of emotions helps explain our supposed ability to identify emotions across cultures by suggesting a smaller number of basic innate emotions shared by all humans - typically happiness, anger, sadness, fear, surprise, and disgust.

One of the common criticisms of BET is its limited amount of emotions and the fact that the association may vary with culture [15]. Some appraisal theory authors suggest the existence of an almost infinite number of emotional states, for which basic emotions would represent only a few. Theorists of this approach argue that the boundaries between emotions are not as clear as the BET suggests. For a subject to experience a specific emotion, it is necessary to evaluate internal and external factors. They also argue that culture is an important factor that influences how we appraise the situation, which ultimately suggests that culture interferes with our emotional experiences [9] [16].

More recent studies have used computational and AI techniques [5] [4] to map the number of emotions in large volumes of data from different modalities (facial, body, and voice expression). These researches were consolidated in the Semantic Space Theory proposal [17], in which the authors cite that emotions are high dimensional, which involves more than 25 different types of emotions, each with an associated response profile. They also pointed out that the boundaries between the categories of emotions are not discrete, with no clear borders, and that the emotional response is systematically blended.

The constructivist theory [13] [1], in turn, rejects the hypothesis that there are well-defined basic and universal emotions, nor that there is a specific biological marker for each emotion. Emotions are a combination of different psychological dimensions. Valence and arousal are primary in emotional experience. People make interpretations about affect from these components, which gives rise to significant individual and cultural variation in the experience of emotions.
The constructivist theory also relies on language and words as necessary components for understanding and perceiving emotions. Some models argues that emotional concepts - such as "anger" or "happiness" are learned inferences about the meaning of physiological processes related to body's homeostasis rather than originated by a dedicated brain structure [14].

Cultural and language adaptations are important, as the subjective values assigned vary across cultures. However, the species has hardwired circuits that indexes and evokes certain types of responses. In this sense, we argue that some responses may be universal, although the subjective value attributed depends on what is relevant to the individual. The role of culture and language seems to account for the existing variations.

This does not mean that we do not feel different sensations without language, but the words help us to group different experiences and divide this continuum into discrete emotions [2] [12]. While different languages share many words for emotion, the meaning could be different among them. An study [10] examined about 2500 languages to determine the degree of similarity of 24 emotion terms across cultures. They found low levels of similarity and high variability in the meaning of emotion terms across cultures. For example, "anxiety" was close to "fear" in Tai-Kadai lanuages while the same emotion word was more telated to "grief" and "regrets" in Austroasiatic languages.

It is important to note that there are even words for emotions in one language without a direct counterpart in another language. For example, the word "saudade" in Portuguese, which refers to a state of deep sadness and melancholy caused by the absence of someone, has no direct translation into English. The same happens with the concept amae in Japanese and Schadenfreude in German, which do not have English counterparts.

### B. Culture and Language in Emotions

Those evidences leads us to argue the importance of thinking about language and culture when planing to use emotional approach in HCI. In a recent study [3] published by the authors in the field of affective computing it was necessary to perform a cross-cultural adaptation for our language and culture. We adapted the framework GoEmotion [6] to Portuguese to develop affective AI projects in Brazil. This adaptation [3] was not only based on direct translation, but involved a group of reviewers (psychologists, neuroscientists, linguists and computer scientists) analyzing the definitions of each category in the original language (English) to determine whether they were applied to portuguese.

The result of this process involved some adaptations: the reviewers suggested changing the emotional category cuidado, translated from caring to compaixão, as it is a more broad and blended emotion in the Portuguese language. There was also a consensus among researchers to add saudade and inveja emotions to the list as they are prevalent to portuguese.

## III. FINAL DISCUSSION

Most technologies and interfaces are designed to be global, but often do not take into account local characteristics. Emotional aspects are being incorporated into HCI and affective computing, so it is important to discuss whether the same emotional framework can be implemented globally regardless of culture.

In this paper, we discussed the main theories of emotions, such as Basic Emotions Theory (BET), constructivist theory, and appraisal theory. We argue that we must consider that some cultural adaptations are important, as the subjective values attributed to the experience vary between cultures. Language and culture account for variation in emotional expression and categorization. For this reason, we should understand language and culture as important criteria for the study of emotions in HCI.

## REFERENCES

[1] LisaFeldmanBarrett.2016.Thetheoryofconstructedemotion: anactiveinferenceaccountofinteroceptionandcategorization.SocialCognitiveand Affective Neuroscience (oct 2016), nsw154. https://doi.org/10.1093/scan/nsw154

[2] Lisa Feldman Barrett, Kristen A. Lindquist, and Maria Gendron. 2007. Language as context for the perception of emotion. Trends in Cognitive Sciences 11, 8 (aug 2007), 327–332. https://doi.org/10.1016/j.tics.2007.06.003

[3] Diogo Cortiz, Jefferson O. Silva, Newton Calegari, Ana Luísa Freitas, Ana Angélica Soares, Carolina Botelho, Gabriel Gaudencio Rêgo, Waldir Sampaio, and Paulo Sergio Boggio. 2021. A Weakly Supervised Dataset of Fine-Grained Emotions in Portuguese. In Anais do XIII Simpósio Brasileiro deTecnologiadaInformaçãoedaLinguagemHumana(STIL2021). SociedadeBrasileiradeComputação,73–81. https://doi.org/10.5753/stil.2021.17786

[4] Alan S. Cowen, Xia Fang, Disa Sauter, and Dacher Keltner. 2020. What music makes us feel: At least 13 dimensions organize subjective experiences associated with music across different cultures. Proceedings of the National Academy of Sciences 117, 4 (jan 2020), 1924–1934. https://doi.org/10.1073/pnas.1910704117

[5] Alan S. Cowen and Dacher Keltner. 2017. Self-report captures 27 distinct categories of emotion bridged by continuous gradients. Proceedings of the National Academy of Sciences 114, 38 (sep 2017), E7900–E7909. https://doi.org/10.1073/pnas.1702247114

[6] Dorottya Demszky, Dana Movshovitz-Attias, Jeongwoo Ko, Alan Cowen, Gaurav Nemade, and Sujith Ravi. 2020. GoEmotions: A Dataset of Fine-Grained Emotions. In Proceedings of the 58th Annual Meeting of the Association for Computational Linguistics. Association for Computational Linguistics, Stroudsburg, PA, USA, 4040–4054. https://doi.org/10.18653/v1/2020.acl-main.372

[7] Paul Ekman. 1992. An argument for basic emotions. Cognition and Emotion 6, 3-4 (may 1992), 169–200. https://doi.org/10.1080/02699939208411068

[8] Paul Ekman. 1993. Facial expression and emotion. American Psychologist 48, 4 (1993), 384–392. https://doi.org/10.1037/0003-066X.48.4.384

[9] Phoebe C. Ellsworth. 2013. Appraisal Theory: Old and New Questions. Emotion Review 5, 2 (apr 2013), 125–131. https://doi.org/10.1177/1754073912463617

[10] Joshua Conrad Jackson, Joseph Watts, Teague R. Henry, Johann-Mattis List, Robert Forkel, Peter J. Mucha, Simon J. Greenhill, Russell D. Gray, and Kristen A. Lindquist. 2019. Emotion semantics show both cultural variation and universal structure. Science 366, 6472 (dec 2019), 1517–1522. https://doi.org/10.1126/science.aaw8160

[11] Magda Kowalska and Monika Wróbel. 2017. Basic Emotions. In Encyclopedia of Personality and Individual

Differences. Springer International Publishing, Cham, 1–6. https://doi.org/10.1007/978-3-319-28099-8_495-1

[12] Kristen A. Lindquist. 2009. Language is Powerful. Emotion Review 1, 1 (jan 2009), 16–18. https://doi.org/10.1177/1754073908097177

[13] Kristen A. Lindquist. 2013. Emotions Emerge from More Basic Psychological Ingredients: A Modern Psychological Constructionist Model. Emotion Review 5, 4 (oct 2013), 356–368. https://doi.org/10.1177/1754073913489750

[14] Kristen A. Lindquist, Tor D. Wager, Hedy Kober, Eliza Bliss-Moreau, and Lisa Feldman Barrett. 2012. The brain basis of emotion: A meta-analytic review. Behavioral and Brain Sciences 35, 3 (jun 2012), 121–143. https://doi.org/10.1017/S0140525X11000446

[15]JamesA.Russell.1994.Isthereuniversalrecognitionofemotionfromfacialexpression?Areviewofthecross-culturalstudies.PsychologicalBulletin 115, 1 (1994), 102–141. https://doi.org/10.1037/0033-2909.115.1.102

[16] K R Scherer. 1999. Appraisal Theory. In Handbook of cognition and emotion.

[17] Cowe, Allan. Keltner, Dacher. Semantic Space Theory: A Computational Approach to Emotion. Trends in Cognitive Science. 2021